\documentclass[reprint,amsmath,amssymb,aps,prl,superscriptaddress]{revtex4-1}
\usepackage{CJK}
\usepackage{graphicx}% Include figure files
\usepackage{dcolumn}% Align table columns on decimal point
\usepackage{bm}% bold math
\emergencystretch=\maxdimen
\begin{document}
\begin{CJK*}{UTF8}{gbsn}

\preprint{APS/123-QED}

\title{Shape Deformation and Drag Variation of a Coupled Rigid-flexible System in a Flowing Soap Film}

\author{Song Gao (高颂)}
\affiliation{State Key Laboratory of Ocean Engineering, School of Naval Architecture, Ocean and Civil Engineering, Shanghai Jiao Tong University, Shanghai 200240, China}
\affiliation{Department of Mechanical Engineering, Northwestern University, Evanston, IL 60208, USA}
\author{Song Pan (潘松)}
\author{Huaicheng Wang (王怀成)}
\affiliation{State Key Laboratory of Ocean Engineering, School of Naval Architecture, Ocean and Civil Engineering, Shanghai Jiao Tong University, Shanghai 200240, China}
\author{Xinliang Tian (田新亮)}
\email{Corresponding author: tianxinliang@sjtu.edu.cn}
\affiliation{State Key Laboratory of Ocean Engineering, School of Naval Architecture, Ocean and Civil Engineering, Shanghai Jiao Tong University, Shanghai 200240, China}
\affiliation{Shanghai Jiao Tong University Yazhou Bay Institute of Deepsea Technology, Sanya 572000, China}

\date{\today}

\begin{abstract}
We experimentally study the flow past a rigid plate with an attached closed filament acting as a deformable afterbody in the soap film. The complex fluid-structure interactions due to its deformable shape and corresponding dynamics are studied. We find the shape of the afterbody is determined by the filament length and flow velocity. A significant drag reduction of up to 9.0\% is achieved by adjusting the filament length. We analyze the drag mechanism by characterizing the deformable afterbody shape and wake properties. Our experiment and modeling suggest that such favorable flow control and drag reduction are expected to occur over a specific flow speed regime when the flexible afterbody is suitably added.
\end{abstract}

\maketitle
\end{CJK*}

Flow past a bluff body is encountered in numerous natural and industrial scenarios. Shape of the body, serving as the boundary for surrounding flow, dominates the fluid force and wake dynamics. Various flow controls \cite{choi2008control}, including drag reduction, lift enhancement and vibration suppression, are achieved effectively by adjusting the shape of the bluff body, e.g., adding dimples \cite{bearman1993control} or a splitter plate \cite{roshko1961experiments, bearman1965investigation, anderson1997effects}.   For aquatic animals, hydrodynamic performance is enhanced by the structural and morphological components of their body\cite{fish2006passive}, e.g., riblets on shark skin \cite{bechert1989viscous} and bumps on whale flippers \cite{van2008bumps}. The bioinspired drag-reducing surfaces/garments have shown efficiency in improving athletes' performance \cite{oeffner2012hydrodynamic}. However, these \textit{passive} control methods requires structure or surface modifications on the rigid body.  Flying birds tend to show elegant and deft ways to increase speed and reduce energy consumption. Their feathers self-adapt during flight, which is beneficial for aerodynamic performance \cite{favier2009passive,niu2011drag,bagheri2012spontaneous}. The shape self-adaptation under flow, also referred to as \textit{reconfiguration} \cite{gosselin2010drag,leclercq2018reconfiguration}, to reduce drag occurs universally in the botanic world as plants seek to facilitate their flexibility to bend, fold and twist when subjected to fluid, both in water \cite{nepf2012flow} and in air \cite{de2008effects,cummins2018separated}. As a result, drag scales up more slowly than classical the drag$\sim$velocity square law for rigid bodys \cite{vogel1989drag,alben2002drag,shelley2011flapping}. However, it seems impossible to make the shape or structure of cars or airplanes compliantly accommodate fluids in the real world. This raises a question of whether adding a suitably flexible coating to a rigid body can import some favorable flow control features and improve its aero/hydrodynamic performance.

In this Letter, we experimentally investigate two-dimensional (2D) flow past a rigid flat plate attached with a closed filament of negligible weight. The compliant filament passively controls the flow over the plate and yields a surprisingly reduced drag. This deformable afterbody can be easily installed/removed and even adjusted in size, making such a flow control method possible and applicable in practical situations.

 \begin{figure*}[htbp]
 \includegraphics[scale=0.78]{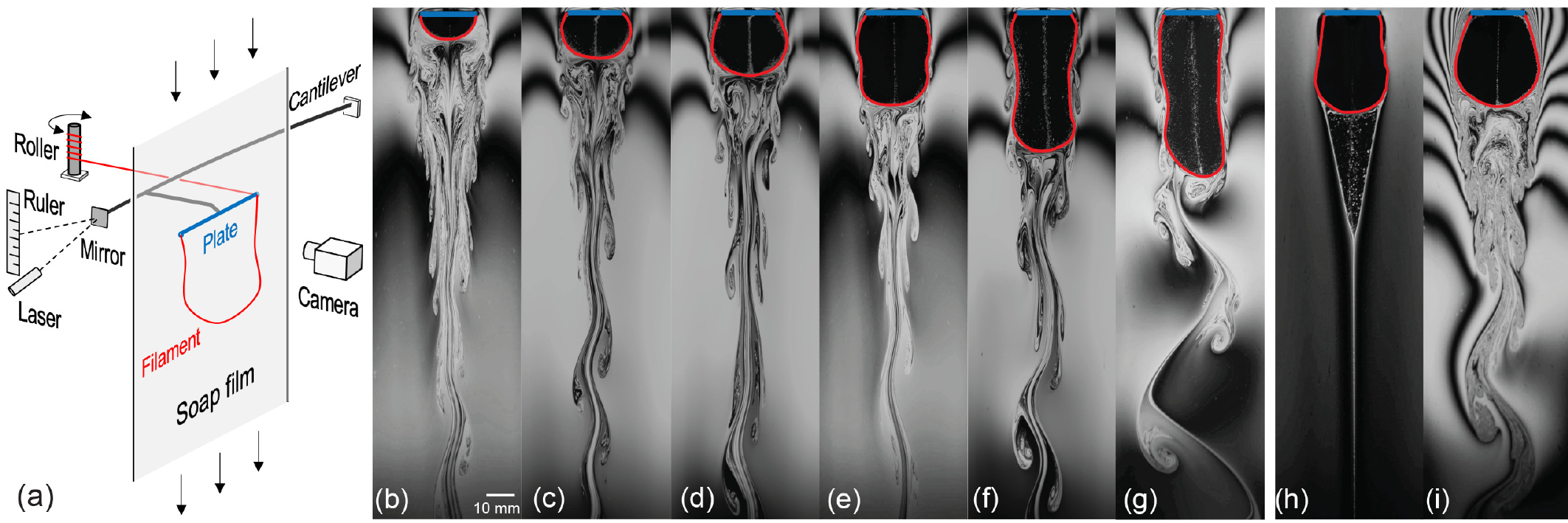}
 \caption{(a) Schematic view of the experiment: test section (not to scale). The plate and the filament are highlighted in blue and red, respectively. Typical flow features at $U=1.45$ m/s: (b) plate-like regime, $\Lambda=1.56$; (c,d) cylinder-like regime, $\Lambda=2.81$ and $\Lambda=3.47$; (e) slender shape regime, $\Lambda=4.25$; (f) rolling vortex regime, $\Lambda=6.46$ and (g) flapping regime, $\Lambda=6.95$. The other two cases (h) $\Lambda=4.26$ at $U_1=0.97$ m/s and (i) $\Lambda=4.30$ at $U_7=1.90$ m/s are comparable with (e) to show the influence of the flow velocity $U$.}
 \label{fig1}
 \end{figure*}

\textit{Experiments.}---Soap film, which functions as a 2D flow tunnel, provides convenience for resembling 2D hydrodynamics in several aspects \cite{rivera1998turbulence,zhang2000flexible,rutgers2001conducting,roushan2005structure,jung2006dynamics,jia2008passive,schnipper2009vortex}. In our vertically flowing soap film
(soapy water density $\rho=1.065$ g/cm$^{3}$), a rigid rod (length $L_p=20$ mm; diameter 0.45 mm) serves as the flat plate placed normal to the incoming flow, and a closed flexible filament
(length embedded in the soap film $L_d=$ 20--140 mm; diameter 12 $\mu$m; bending stiffness 3.43$\times 10^{-3}$ g$\cdot$cm$^{3}$/s$^{2}$; linear density 1.96$\times 10^{-6}$ g/cm) attached downstream behaves as the deformable afterbody. We use a dimensionless scale $\Lambda=L_d/L_p$ to describe the geometry of this rigid-flexible coupled system, where $\Lambda=1$ refers to the situation with no afterbody attached. The upper bound of $\Lambda$ is approximately 7, beyond which a stable and long-lasting flow cannot be achieved. In the parallel test section, a uniform velocity profile and constant film thickness are approached over 70\% of the span about the midline. The tunnel is wide enough (tunnel width, 110 mm) that no obvious blockage is observed. The wake patterns are visualized by an interference technique using the monochromatic light of a low-pressure sodium lamp. The fluid drag acting on the plate is obtained by measuring the vertical displacement of the supporting cantilever \cite{alben2002drag}. This method is proven to be both statically and dynamically reliable \cite{jia2009response}. During the experiment, $L_d$ can be modified gently and continuously without suspending or even disturbing the flowing film \cite{experimentcon}, as shown in Fig.~\ref{fig1}(a). Seven different flow velocities $U$ through the range 0.97--1.90 m/s are tested, and approximately 150 to 200 sets of measurements under different $\Lambda$ are conducted per $U$. Each drag data point is time-averaged over 30 seconds. The filament is wetted by the fluid and constrained in the plane of film always, and the deformable afterbody appears to bend only and without measurable change in length. The filament is much thicker than the soap film (thickness $f=$ 1--3 $\mu$m); thus, no obvious inside-outside fluid exchange is observed either. The Marangoni wave speed \cite{couder1989hydrodynamics} is considerably larger than the flow speeds, and thus, no significant effects of compressibility are considered. The kinematic viscosity of flowing soap film is $\nu=0.07\ \text{cm}^2/\text{s}$.  The Reynolds number Re $=L_pU/\nu$ is approximately 2700-5500.

\begin{figure*}[t]
\includegraphics[scale=0.94]{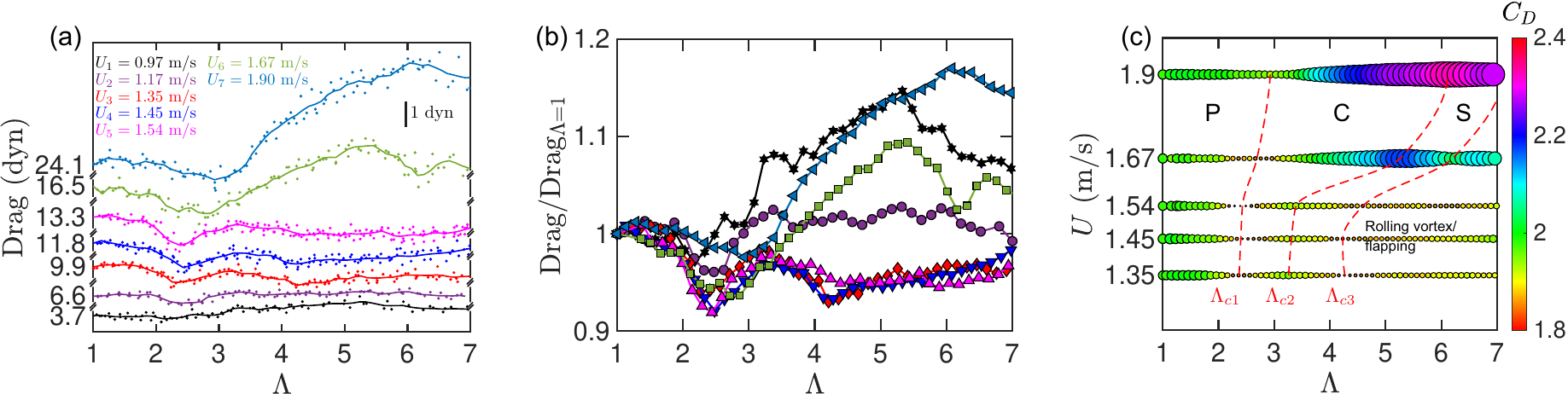}
\caption{ (a) Fluid drag of the whole system as a function of $\Lambda$ at different $U$. Solid lines denote the average fitting value of the measured drag data points, shown as dots. (b) Fluid drag is normalized by the drag at $\Lambda=1$. (c) Flow regimes in the $U-\Lambda$ map. Thresholds are defined by the local minimum or maximum drag coefficient $C_D$. Color and size of the symbols both indicate the magnitude of $C_D$. $U_1$ and $U_2$ are not shown due to very small drag variation and different physics.}
\label{fig2}
\end{figure*}

\textit{Flow pattern.}---The typical flow features at $U=1.45$ m/s are illustrated in Fig.~\ref{fig1}(b-g). As shown in Fig.~\ref{fig1}(b), the flow at $\Lambda=1.56$ is almost the same as that for a 2D flat plate, i.e., the flow separates at the plate edges, and a pair of counter-rotating rings is formed at the rear of the plate. A very short deformable body occupies only some area of the wake, imposing no obvious influence on the ambient flow. Therefore, the first regime in the stationary state is denoted as `plate-like' (P) regime. As $\Lambda$ increases, fluid plumps up the deformable afterbody at the edges of the plate [Fig.~\ref{fig1}(c,d)]. The incoming flow passes along the filament, and separation occurs on the rear part of the deformable afterbody rather than at the plate edges. The significant separation delay observed here resembles the flow past a 2D cylinder \cite{williamson1996vortex}, indicating the onset of the `cylinder-like' (C) regime. The width of the deformable body in the C regime increases as $\Lambda$ increases until the beginning of the `slender shape' (S) regime [Fig.~\ref{fig1}(e)]. In the S regime, the middle section of the afterbody is squeezed by the outside fluid and becomes narrower. Moreover, the longer the filament is, the narrower the profile is. In these aforementioned regimes, the afterbody appears stationary and reflectional symmetry about the midline, behaving as a rigid body. Beyond the S regime, flow enters the `rolling vortex' regime, where the afterbody shape resembles that in the S regime but the filament traps significant vortex to roll along the two sides [Fig.~\ref{fig1}(f)]. Finally, when the filament is long enough that it flaps, the last flapping regime is achieved [Fig.~\ref{fig1}(g)]. 

\textit{Drag variation.}---We further investigate the fluid drag acting on the coupled system to better understand the transitions between different flow regimes. The fluid drag is normalized by the drag of the bare plate ($\Lambda=1$) to compare drag variation at different $U$, respectively [Fig.~\ref{fig2}(b)]. Additionally, the drag coefficient $C_D=\text{Drag}/(\rho U^2L_pf/2)$ is introduced to scale the fluid drag with the flow velocity. It is noted that $f$ of the soap film is dependent on $U$ as $f\propto U^{0.75}$ \cite{sane2018surface}. After taking this into account, we find all $C_D$ gather together at approximately 2.0-2.1 at $\Lambda=1$ [Fig.~\ref{fig2}(c)], which agrees with the reported results  \cite{anatol1955wake,munson2013fluid}. 

When flow velocity exceeds $U_2=1.17$ m/s, the five drag curves show a similar tendency as $\Lambda$ varies, and we find the drag variation is closely related with the transitions between flow regimes reported above. The normalized drag decreases until $\Lambda$ reaches the first threshold at $\Lambda_{c1}$, where the onset of the C regime results in a local minimum drag. Then, the drag increases due to the growth of afterbody width in the C regime until the second threshold at $\Lambda_{c2}$. After entering the S regime, the system benefits from its narrow shape and displays the second decrease in total drag. However, when the S regime ends at $\Lambda_{c3}$, the rolling vortex on the sides and the flapping of the deformable afterbody reverses the trend, leading to a general increasing trend of drag. It is hard to explore the drag in these two regimes since the afterbody arbitrarily changes its shape or flaps strongly, making the soap film susceptible to rupture and large measurement fluctuations. The heavier drag burden suffered by the whole system distinguishes these two regimes from previous three regimes, yet the transition between these two regimes is unstable and still not well understood. The reasons that the drag variation is not obvious at $U=0.97$ and 1.17 m/s are that, drag is extremely small at such small $U$ so that the drag variation is even more minuscule; different physics occurs (will be explained hereinafter). It is noteworthy that the curves of normalized drag at $U=1.35$, 1.45 and 1.54 m/s collapse, and the critical $\Lambda$ values are very close as well ($\Lambda_{c1}\approx2.3$, $\Lambda_{c2}\approx3.3$ and $\Lambda_{c3}\approx4.2$). Moreover, the most dramatic drag reduction of approximately 9.0\% (compared with the bare plate drag) is observed at $\Lambda_{c1}$ in this speed zone. Thus this range of velocity is referred to as the `favorable drag zone' (FDZ) in the following parts. Out of the FDZ, the system behaves in different ways. First, as $U$ leaves further away from the FDZ , the normalized drag curves deviates more (see black stars at $U_1=0.97$ m/s and blue left-pointing triangles at $U_7=1.90$ m/s), which means using bare plate drag to normalize drag regardless of the deformable shape of the afterbody has inherent limitations, especially when $U$ is small/large enough to affect its shape in different ways; second, the transition from C to S regime (denoted by $\Lambda_{c2}$) is hysteretic for large $U$, i.e., the C regime lasts significantly longer at higher speeds. This is confirmed by the deformable afterbody shape at different $U$ given the same $\Lambda$. The afterbody is easily squeezed at small $U$ such that the section of filament near plate is even embedded in the wake [Fig.~\ref{fig1}(h)], while it grows even wider at larger $U$ [Fig.~\ref{fig1}(i)]. These phenomena suggest that the shape of the deformable afterbody and its suffered drag are closely related, and both of them are affected by the filament length and flow velocity.  Therefore, the dependence of afterbody shape on  $\Lambda$ and $U$ is investigated. 

\begin{figure*}[t]
\includegraphics[scale=0.82]{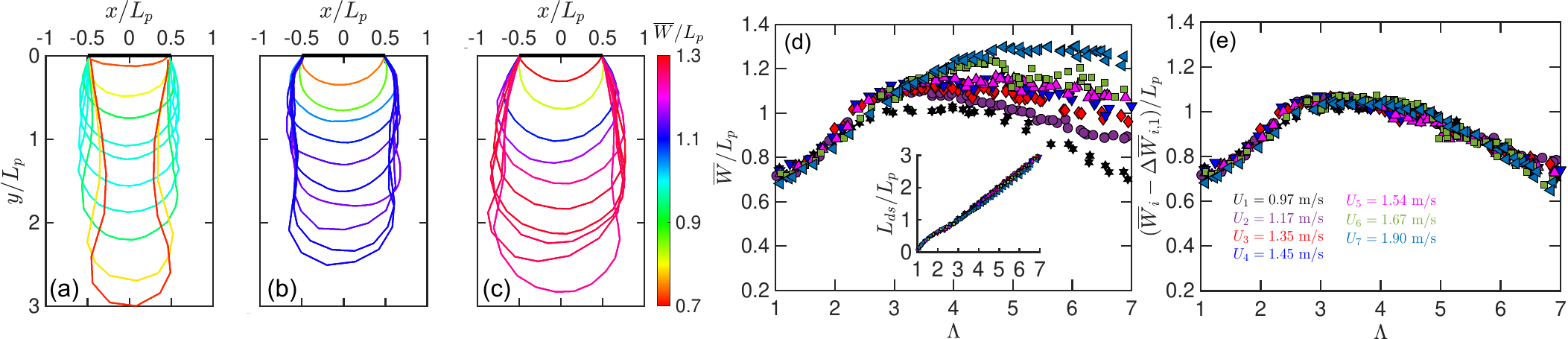}
\caption{Shape reconfiguration of the deformable afterbody as $\Lambda$ varies at (a) $U_1=0.97$ m/s, (b) $U_5=1.54$ m/s and (c) $U_7=1.90$ m/s. $x$ and $y$ are the spanwise and streamwise coordinates, respectively. The color of the filament indicates the dimensionless average width of the afterbody, $\overline{W}/L_p$. The afterbody are generally wider at higher $U$. (d) Variations of $\overline{W}/L_p$ and $L_{ds}/L_p$ versus $\Lambda$ at different $U$. (e) Rescaling of $\overline{W}$ using the model expressed by Eq.~\ref{eq1}.}
\label{fig3}
\end{figure*}

\textit{Shape deformation.}---The shapes of the deformable afterbody at different $\Lambda$ are shown in Fig.~\ref{fig3}(a-c). The average width of afterbody $\overline{W}$ is calculated as the ratio of the enclosed area over the maximum afterbody length $L_{ds}$ (from the plate to the farthest downstream point). $\overline{W}$ shows the same tendency for all velocities. As $\Lambda$ increases, $\overline{W}$ first increases, corresponding to the transition from the P to the C regime and the growth of afterbody. Then, $\overline{W}$ decreases as the flow transits to the S regime. Moreover, a larger velocity gives an overall larger $\overline{W}$ at the same $\Lambda$, which agrees with the phenomena observed in Fig.~\ref{fig1}(e,h,i) and Fig.~\ref{fig3}(a,b,c). It is observed in Fig.~\ref{fig3}(d) that all $\overline{W}/L_p$ curves collapse for different $U$ when $\Lambda\le2.3$. It is reasonable that in the P regime, the filament is short and trapped in the wake so that its shape is not significantly influenced by the flow out of the separated free-shear layer. Beyond P regime, the deformable afterbody tends to grow more in width at larger $U$. Such a difference also becomes more pronounced at large $\Lambda$ side. $U$ and $\Lambda$ together determine the average width of the deformable afterbody. Since there are no inside-outside fluid interactions and the inside velocity $u$ is at least one magnitude less than that outside, the significant velocity difference causes a fluid pressure difference $0.5\rho(U^2-u^2)\sim0.5\rho U^2$ acting on the two sides of the deformable afterbody. On the other hand, the afterbody length increases linearly with $\Lambda$ beyond $\Lambda_{c1}$, i.e., $L_{ds}\sim c(\Lambda-\Lambda_{c1})$, determining the real force acting on the sides. Such force broadens or narrows the afterbody in width. Given any two flow velocities $U_i$ and $U_j$, the corresponding width difference $\Delta\overline{W}_{i,j}=\overline{W}_i-\overline{W}_j$ beyond $\Lambda_{c1}$ can be written as
\begin{equation}
  \Delta\overline{W}_{i,j}\sim0.5kc\rho(U_i^2-U_j^2)(\Lambda-\Lambda_{c1})
\label{eq1}
\end{equation}
where $k$ is a constant fitting parameter with the unit of m$\cdot$s$^2$/kg that takes the filament bending stiffness, afterbody mass, etc. into consideration. This form is self-consistent in that for the same $U$, there is no width difference; for $\Lambda=2.3$, no width difference exists either. If we eliminate the width difference between $U_i$ and $U_1$ (black star) by $0.5kc\rho(U_i^2-U_1^2)(\Lambda-\Lambda_{c1})$, all curves remarkably collapse on $\overline{W}_1$ [Fig.~\ref{fig3}(e)]. This model is capable of explaining the physics underlying the width difference between different $U$. Although the method to fully resolve the afterbody shape and curvature at any given ($\Lambda$, $U$) has yet to be determined, it is reasonable that $\overline{W}$ could be represented by a function of $\Lambda$ and $U$.  Our model based on the pressure difference is helpful to explain the phenomena described previously. First, the delayed transition from C to S regime: the larger $U$ is, the smaller the outside pressure is. Thus, the afterbody tends to grow more in width, and the C regime lasts longer. Second, the different physics at $U_1$ and $U_2$: the afterbody shape is easily depressed at small $U$, so that the separation points return to the plate edges [Fig.~\ref{fig1}(h)], making the drag increase sharply [Fig.~\ref{fig2}(b)]. Since the fluid drag varies in a different way, $U_1$ and $U_2$ are not taken into discussion in the following part.

\begin{figure}[b]
\includegraphics[scale=0.88]{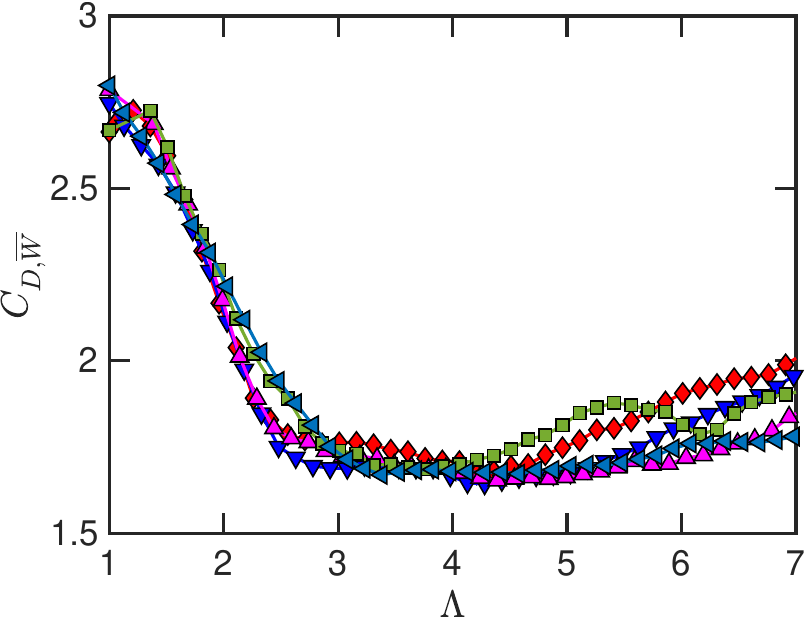}
\caption{Drag is rescaled after accounting for the shape deformation. Drag coefficient $C_{D,\overline{W}}$, normalized using $\overline{W}$, is plotted as a function of $\Lambda$. }
\label{fig4}
\end{figure}

\textit{Drag scaling based on the afterbody shape.}---Since the afterbody width $\overline{W}$ is also a function of $\Lambda$ and $U$, using $\overline{W}$ as the characteristic length to normalize the drag gives more physical insights on the drag scaling. As shown in Fig.~\ref{fig4}, all $C_{D,\overline{W}}$ show similar trends and approximately coincide in the P, C and S regimes. It is noted that the drag coefficient curves of the cases above the FDZ ($U_6$, green squares; $U_7$, blue left-pointing triangles) get closer to the cases in the FDZ than the previous scaling based on $L_p$  [Fig.~\ref{fig2}(b)], suggesting the reconfiguration of the afterbody causes the discrepancy given by the previous scaling without considering the shape deformation. More importantly, similar physics, including shape deformation and drag variation, are reasonably expected to occur at even larger Re regime than the FDZ, though the drag reduction is not that significant. Thus, for similar moving rigid objects in the similar Re regime, with the addition of a suitably flexible coating, such drag reduction will take place as well. One should note that a larger discrepancy is observed even for FDZ cases beyond the S regime ($\Lambda>4.2$). The possible reason is that the contribution from form drag and skin friction drag to the total drag changes. In the previous study on bluff body flow in the soap film \cite{alben2002drag}, the form drag predominates due to the small upper bound of the estimated skin frictional drag \cite{bearman1967vortex,batchelor2000introduction}. However, for our coupled system, form drag is no more the exclusively dominating factor, especially in the rolling vortex and flapping regimes (One observation is that $\overline{W}$ keeps decreasing in these two regimes [Fig.~\ref{fig3}(d)] but the whole system suffers a heavier drag burden [Fig.~\ref{fig2}(b)]). A drag scaling law over all the regimes can be proposed if the drag contribution can be better understood. 

In summary, we investigate the 2D flow past a coupled rigid-flexible system in a flowing soap film, and focus on the shape of the deformable afterbody and the fluid drag acting on the whole system. The flexible afterbody keeps stationary when its length is short, in contrast to the common sense that flexible loops should flap in flowing fluids \cite{jung2006dynamics,shoele2010flow,kim2012flexible}, but at the same time, its shape reconfigures according to the fluid, which occurs ubiquitously in the biological world. By exploring the underlying physics of this new class of fluid-structure interactions, we expect such shape reconfiguration and flow features to be repeated in a similar and even larger Re regime for bluff bodies. Due to the negligible additional weight and convenient installation/disposal of the flexible afterbody, controlling the flow around a bluff body and reducing its suffered drag in a simple way are possible, which may benefit the performance of athletes, racing cars, diving submarines, etc., and possibly inspire novel designs in many areas.

The authors acknowledge Yufeng Kou and Xia Wu for assisting with the experiment. This work is supported by the Natural Science Foundation of Shanghai (Grant No. 19ZR1426300) and National Natural Science Foundation of China (Grant No. 11632011).

\nocite{*}

%\bibliographystyle{apsrev4-1}
%\bibliography{references}

\providecommand{\noopsort}[1]{}\providecommand{\singleletter}[1]{#1}%
\begin{thebibliography}{40}%
\makeatletter
\providecommand \@ifxundefined [1]{%
 \@ifx{#1\undefined}
}%
\providecommand \@ifnum [1]{%
 \ifnum #1\expandafter \@firstoftwo
 \else \expandafter \@secondoftwo
 \fi
}%
\providecommand \@ifx [1]{%
 \ifx #1\expandafter \@firstoftwo
 \else \expandafter \@secondoftwo
 \fi
}%
\providecommand \natexlab [1]{#1}%
\providecommand \enquote  [1]{``#1''}%
\providecommand \bibnamefont  [1]{#1}%
\providecommand \bibfnamefont [1]{#1}%
\providecommand \citenamefont [1]{#1}%
\providecommand \href@noop [0]{\@secondoftwo}%
\providecommand \href [0]{\begingroup \@sanitize@url \@href}%
\providecommand \@href[1]{\@@startlink{#1}\@@href}%
\providecommand \@@href[1]{\endgroup#1\@@endlink}%
\providecommand \@sanitize@url [0]{\catcode `\\12\catcode `\$12\catcode
  `\&12\catcode `\#12\catcode `\^12\catcode `\_12\catcode `\%12\relax}%
\providecommand \@@startlink[1]{}%
\providecommand \@@endlink[0]{}%
\providecommand \url  [0]{\begingroup\@sanitize@url \@url }%
\providecommand \@url [1]{\endgroup\@href {#1}{\urlprefix }}%
\providecommand \urlprefix  [0]{URL }%
\providecommand \Eprint [0]{\href }%
\providecommand \doibase [0]{http://dx.doi.org/}%
\providecommand \selectlanguage [0]{\@gobble}%
\providecommand \bibinfo  [0]{\@secondoftwo}%
\providecommand \bibfield  [0]{\@secondoftwo}%
\providecommand \translation [1]{[#1]}%
\providecommand \BibitemOpen [0]{}%
\providecommand \bibitemStop [0]{}%
\providecommand \bibitemNoStop [0]{.\EOS\space}%
\providecommand \EOS [0]{\spacefactor3000\relax}%
\providecommand \BibitemShut  [1]{\csname bibitem#1\endcsname}%
\let\auto@bib@innerbib\@empty
%</preamble>
\bibitem [{\citenamefont {Bearman}\ and\ \citenamefont
  {Harvey}(1993)}]{bearman1993control}%
  \BibitemOpen
  \bibfield  {author} {\bibinfo {author} {\bibfnamefont {P.~W.}\ \bibnamefont
  {Bearman}}\ and\ \bibinfo {author} {\bibfnamefont {J.~K.}\ \bibnamefont
  {Harvey}},\ }\href@noop {} {\bibfield  {journal} {\bibinfo  {journal} {AIAA
  J.}\ }\textbf {\bibinfo {volume} {31}},\ \bibinfo {pages} {1753} (\bibinfo
  {year} {1993})}\BibitemShut {NoStop}%
\bibitem [{\citenamefont {Roshko}(1961)}]{roshko1961experiments}%
  \BibitemOpen
  \bibfield  {author} {\bibinfo {author} {\bibfnamefont {A.}~\bibnamefont
  {Roshko}},\ }\href@noop {} {\bibfield  {journal} {\bibinfo  {journal} {J.
  Fluid Mech.}\ }\textbf {\bibinfo {volume} {10}},\ \bibinfo {pages} {345}
  (\bibinfo {year} {1961})}\BibitemShut {NoStop}%
\bibitem [{\citenamefont {Bearman}(1965)}]{bearman1965investigation}%
  \BibitemOpen
  \bibfield  {author} {\bibinfo {author} {\bibfnamefont {P.~W.}\ \bibnamefont
  {Bearman}},\ }\href@noop {} {\bibfield  {journal} {\bibinfo  {journal} {J.
  Fluid Mech.}\ }\textbf {\bibinfo {volume} {21}},\ \bibinfo {pages} {241}
  (\bibinfo {year} {1965})}\BibitemShut {NoStop}%
\bibitem [{\citenamefont {Anderson}\ and\ \citenamefont
  {Szewczyk}(1997)}]{anderson1997effects}%
  \BibitemOpen
  \bibfield  {author} {\bibinfo {author} {\bibfnamefont {E.~A.}\ \bibnamefont
  {Anderson}}\ and\ \bibinfo {author} {\bibfnamefont {A.~A.}\ \bibnamefont
  {Szewczyk}},\ }\href@noop {} {\bibfield  {journal} {\bibinfo  {journal} {Exp.
  Fluids}\ }\textbf {\bibinfo {volume} {23}},\ \bibinfo {pages} {161} (\bibinfo
  {year} {1997})}\BibitemShut {NoStop}%
\bibitem [{\citenamefont {Choi}\ \emph {et~al.}(2008)\citenamefont {Choi},
  \citenamefont {Jeon},\ and\ \citenamefont {Kim}}]{choi2008control}%
  \BibitemOpen
  \bibfield  {author} {\bibinfo {author} {\bibfnamefont {H.}~\bibnamefont
  {Choi}}, \bibinfo {author} {\bibfnamefont {W.~P.}\ \bibnamefont {Jeon}}, \
  and\ \bibinfo {author} {\bibfnamefont {J.}~\bibnamefont {Kim}},\ }\href@noop
  {} {\bibfield  {journal} {\bibinfo  {journal} {Annu. Rev. Fluid Mech.}\
  }\textbf {\bibinfo {volume} {40}},\ \bibinfo {pages} {113} (\bibinfo {year}
  {2008})}\BibitemShut {NoStop}%
\bibitem [{\citenamefont {Fish}\ and\ \citenamefont
  {Lauder}(2006)}]{fish2006passive}%
  \BibitemOpen
  \bibfield  {author} {\bibinfo {author} {\bibfnamefont {F.~E.}\ \bibnamefont
  {Fish}}\ and\ \bibinfo {author} {\bibfnamefont {G.~V.}\ \bibnamefont
  {Lauder}},\ }\href@noop {} {\bibfield  {journal} {\bibinfo  {journal} {Annu.
  Rev. Fluid Mech.}\ }\textbf {\bibinfo {volume} {38}},\ \bibinfo {pages} {193}
  (\bibinfo {year} {2006})}\BibitemShut {NoStop}%
\bibitem [{\citenamefont {Bechert}\ and\ \citenamefont
  {Bartenwerfer}(1989)}]{bechert1989viscous}%
  \BibitemOpen
  \bibfield  {author} {\bibinfo {author} {\bibfnamefont {D.~W.}\ \bibnamefont
  {Bechert}}\ and\ \bibinfo {author} {\bibfnamefont {M.}~\bibnamefont
  {Bartenwerfer}},\ }\href@noop {} {\bibfield  {journal} {\bibinfo  {journal}
  {J. Fluid Mech.}\ }\textbf {\bibinfo {volume} {206}},\ \bibinfo {pages} {105}
  (\bibinfo {year} {1989})}\BibitemShut {NoStop}%
\bibitem [{\citenamefont {Van~Nierop}\ \emph {et~al.}(2008)\citenamefont
  {Van~Nierop}, \citenamefont {Alben},\ and\ \citenamefont
  {Brenner}}]{van2008bumps}%
  \BibitemOpen
  \bibfield  {author} {\bibinfo {author} {\bibfnamefont {E.~A.}\ \bibnamefont
  {Van~Nierop}}, \bibinfo {author} {\bibfnamefont {S.}~\bibnamefont {Alben}}, \
  and\ \bibinfo {author} {\bibfnamefont {M.~P.}\ \bibnamefont {Brenner}},\
  }\href@noop {} {\bibfield  {journal} {\bibinfo  {journal} {Phys. Rev. Lett.}\
  }\textbf {\bibinfo {volume} {100}},\ \bibinfo {pages} {054502} (\bibinfo
  {year} {2008})}\BibitemShut {NoStop}%
\bibitem [{\citenamefont {Oeffner}\ and\ \citenamefont
  {Lauder}(2012)}]{oeffner2012hydrodynamic}%
  \BibitemOpen
  \bibfield  {author} {\bibinfo {author} {\bibfnamefont {J.}~\bibnamefont
  {Oeffner}}\ and\ \bibinfo {author} {\bibfnamefont {G.~V.}\ \bibnamefont
  {Lauder}},\ }\href@noop {} {\bibfield  {journal} {\bibinfo  {journal} {J.
  Exp. Biol.}\ }\textbf {\bibinfo {volume} {215}},\ \bibinfo {pages} {785}
  (\bibinfo {year} {2012})}\BibitemShut {NoStop}%
\bibitem [{\citenamefont {Favier}\ \emph {et~al.}(2009)\citenamefont {Favier},
  \citenamefont {Dauptain}, \citenamefont {Basso},\ and\ \citenamefont
  {Bottaro}}]{favier2009passive}%
  \BibitemOpen
  \bibfield  {author} {\bibinfo {author} {\bibfnamefont {J.}~\bibnamefont
  {Favier}}, \bibinfo {author} {\bibfnamefont {A.}~\bibnamefont {Dauptain}},
  \bibinfo {author} {\bibfnamefont {D.}~\bibnamefont {Basso}}, \ and\ \bibinfo
  {author} {\bibfnamefont {A.}~\bibnamefont {Bottaro}},\ }\href@noop {}
  {\bibfield  {journal} {\bibinfo  {journal} {J. Fluid Mech.}\ }\textbf
  {\bibinfo {volume} {627}},\ \bibinfo {pages} {451} (\bibinfo {year}
  {2009})}\BibitemShut {NoStop}%
\bibitem [{\citenamefont {Niu}\ and\ \citenamefont {Hu}(2011)}]{niu2011drag}%
  \BibitemOpen
  \bibfield  {author} {\bibinfo {author} {\bibfnamefont {J.}~\bibnamefont
  {Niu}}\ and\ \bibinfo {author} {\bibfnamefont {D.~L.}\ \bibnamefont {Hu}},\
  }\href@noop {} {\bibfield  {journal} {\bibinfo  {journal} {Phys. Fluids}\
  }\textbf {\bibinfo {volume} {23}},\ \bibinfo {pages} {101701} (\bibinfo
  {year} {2011})}\BibitemShut {NoStop}%
\bibitem [{\citenamefont {Bagheri}\ \emph {et~al.}(2012)\citenamefont
  {Bagheri}, \citenamefont {Mazzino},\ and\ \citenamefont
  {Bottaro}}]{bagheri2012spontaneous}%
  \BibitemOpen
  \bibfield  {author} {\bibinfo {author} {\bibfnamefont {S.}~\bibnamefont
  {Bagheri}}, \bibinfo {author} {\bibfnamefont {A.}~\bibnamefont {Mazzino}}, \
  and\ \bibinfo {author} {\bibfnamefont {A.}~\bibnamefont {Bottaro}},\
  }\href@noop {} {\bibfield  {journal} {\bibinfo  {journal} {Phys. Rev. Lett.}\
  }\textbf {\bibinfo {volume} {109}},\ \bibinfo {pages} {154502} (\bibinfo
  {year} {2012})}\BibitemShut {NoStop}%
\bibitem [{\citenamefont {Gosselin}\ \emph {et~al.}(2010)\citenamefont
  {Gosselin}, \citenamefont {De~Langre},\ and\ \citenamefont
  {Machado-Almeida}}]{gosselin2010drag}%
  \BibitemOpen
  \bibfield  {author} {\bibinfo {author} {\bibfnamefont {F.}~\bibnamefont
  {Gosselin}}, \bibinfo {author} {\bibfnamefont {E.}~\bibnamefont {De~Langre}},
  \ and\ \bibinfo {author} {\bibfnamefont {B.~A.}\ \bibnamefont
  {Machado-Almeida}},\ }\href@noop {} {\bibfield  {journal} {\bibinfo
  {journal} {Journal of Fluid Mechanics}\ }\textbf {\bibinfo {volume} {650}},\
  \bibinfo {pages} {319} (\bibinfo {year} {2010})}\BibitemShut {NoStop}%
\bibitem [{\citenamefont {Leclercq}\ and\ \citenamefont
  {de~Langre}(2018)}]{leclercq2018reconfiguration}%
  \BibitemOpen
  \bibfield  {author} {\bibinfo {author} {\bibfnamefont {T.}~\bibnamefont
  {Leclercq}}\ and\ \bibinfo {author} {\bibfnamefont {E.}~\bibnamefont
  {de~Langre}},\ }\href@noop {} {\bibfield  {journal} {\bibinfo  {journal}
  {Journal of Fluid Mechanics}\ }\textbf {\bibinfo {volume} {838}},\ \bibinfo
  {pages} {606} (\bibinfo {year} {2018})}\BibitemShut {NoStop}%
\bibitem [{\citenamefont {Nepf}(2012)}]{nepf2012flow}%
  \BibitemOpen
  \bibfield  {author} {\bibinfo {author} {\bibfnamefont {H.~M.}\ \bibnamefont
  {Nepf}},\ }\href@noop {} {\bibfield  {journal} {\bibinfo  {journal} {Annu.
  Rev. Fluid Mech.}\ }\textbf {\bibinfo {volume} {44}},\ \bibinfo {pages} {123}
  (\bibinfo {year} {2012})}\BibitemShut {NoStop}%
\bibitem [{\citenamefont {de~Langre}(2008)}]{de2008effects}%
  \BibitemOpen
  \bibfield  {author} {\bibinfo {author} {\bibfnamefont {E.}~\bibnamefont
  {de~Langre}},\ }\href@noop {} {\bibfield  {journal} {\bibinfo  {journal}
  {Annu. Rev. Fluid Mech.}\ }\textbf {\bibinfo {volume} {40}},\ \bibinfo
  {pages} {141} (\bibinfo {year} {2008})}\BibitemShut {NoStop}%
\bibitem [{\citenamefont {Cummins}\ \emph {et~al.}(2018)\citenamefont
  {Cummins}, \citenamefont {Seale}, \citenamefont {Macente}, \citenamefont
  {Certini}, \citenamefont {Mastropaolo}, \citenamefont {Viola},\ and\
  \citenamefont {Nakayama}}]{cummins2018separated}%
  \BibitemOpen
  \bibfield  {author} {\bibinfo {author} {\bibfnamefont {C.}~\bibnamefont
  {Cummins}}, \bibinfo {author} {\bibfnamefont {M.}~\bibnamefont {Seale}},
  \bibinfo {author} {\bibfnamefont {A.}~\bibnamefont {Macente}}, \bibinfo
  {author} {\bibfnamefont {D.}~\bibnamefont {Certini}}, \bibinfo {author}
  {\bibfnamefont {E.}~\bibnamefont {Mastropaolo}}, \bibinfo {author}
  {\bibfnamefont {I.~M.}\ \bibnamefont {Viola}}, \ and\ \bibinfo {author}
  {\bibfnamefont {N.}~\bibnamefont {Nakayama}},\ }\href@noop {} {\bibfield
  {journal} {\bibinfo  {journal} {Nature (London)}\ }\textbf {\bibinfo {volume}
  {562}},\ \bibinfo {pages} {414} (\bibinfo {year} {2018})}\BibitemShut
  {NoStop}%
\bibitem [{\citenamefont {Vogel}(1989)}]{vogel1989drag}%
  \BibitemOpen
  \bibfield  {author} {\bibinfo {author} {\bibfnamefont {S.}~\bibnamefont
  {Vogel}},\ }\href@noop {} {\bibfield  {journal} {\bibinfo  {journal} {J. Exp.
  Bot.}\ }\textbf {\bibinfo {volume} {40}},\ \bibinfo {pages} {941} (\bibinfo
  {year} {1989})}\BibitemShut {NoStop}%
\bibitem [{\citenamefont {Alben}\ \emph {et~al.}(2002)\citenamefont {Alben},
  \citenamefont {Shelley},\ and\ \citenamefont {Zhang}}]{alben2002drag}%
  \BibitemOpen
  \bibfield  {author} {\bibinfo {author} {\bibfnamefont {S.}~\bibnamefont
  {Alben}}, \bibinfo {author} {\bibfnamefont {M.}~\bibnamefont {Shelley}}, \
  and\ \bibinfo {author} {\bibfnamefont {J.}~\bibnamefont {Zhang}},\
  }\href@noop {} {\bibfield  {journal} {\bibinfo  {journal} {Nature (London)}\
  }\textbf {\bibinfo {volume} {420}},\ \bibinfo {pages} {479} (\bibinfo {year}
  {2002})}\BibitemShut {NoStop}%
\bibitem [{\citenamefont {Shelley}\ and\ \citenamefont
  {Zhang}(2011)}]{shelley2011flapping}%
  \BibitemOpen
  \bibfield  {author} {\bibinfo {author} {\bibfnamefont {M.}~\bibnamefont
  {Shelley}}\ and\ \bibinfo {author} {\bibfnamefont {J.}~\bibnamefont
  {Zhang}},\ }\href@noop {} {\bibfield  {journal} {\bibinfo  {journal} {Annu.
  Rev. Fluid Mech.}\ }\textbf {\bibinfo {volume} {43}},\ \bibinfo {pages} {449}
  (\bibinfo {year} {2011})}\BibitemShut {NoStop}%
\bibitem [{\citenamefont {Rivera}\ \emph {et~al.}(1998)\citenamefont {Rivera},
  \citenamefont {Vorobieff},\ and\ \citenamefont
  {Ecke}}]{rivera1998turbulence}%
  \BibitemOpen
  \bibfield  {author} {\bibinfo {author} {\bibfnamefont {M.}~\bibnamefont
  {Rivera}}, \bibinfo {author} {\bibfnamefont {P.}~\bibnamefont {Vorobieff}}, \
  and\ \bibinfo {author} {\bibfnamefont {R.~E.}\ \bibnamefont {Ecke}},\
  }\href@noop {} {\bibfield  {journal} {\bibinfo  {journal} {Phys. Rev. Lett.}\
  }\textbf {\bibinfo {volume} {81}},\ \bibinfo {pages} {1417} (\bibinfo {year}
  {1998})}\BibitemShut {NoStop}%
\bibitem [{\citenamefont {Zhang}\ \emph {et~al.}(2000)\citenamefont {Zhang},
  \citenamefont {Childress}, \citenamefont {Libchaber},\ and\ \citenamefont
  {Shelley}}]{zhang2000flexible}%
  \BibitemOpen
  \bibfield  {author} {\bibinfo {author} {\bibfnamefont {J.}~\bibnamefont
  {Zhang}}, \bibinfo {author} {\bibfnamefont {S.}~\bibnamefont {Childress}},
  \bibinfo {author} {\bibfnamefont {A.}~\bibnamefont {Libchaber}}, \ and\
  \bibinfo {author} {\bibfnamefont {M.}~\bibnamefont {Shelley}},\ }\href@noop
  {} {\bibfield  {journal} {\bibinfo  {journal} {Nature (London)}\ }\textbf
  {\bibinfo {volume} {408}},\ \bibinfo {pages} {835} (\bibinfo {year}
  {2000})}\BibitemShut {NoStop}%
\bibitem [{\citenamefont {Roushan}\ and\ \citenamefont
  {Wu}(2005)}]{roushan2005structure}%
  \BibitemOpen
  \bibfield  {author} {\bibinfo {author} {\bibfnamefont {P.}~\bibnamefont
  {Roushan}}\ and\ \bibinfo {author} {\bibfnamefont {X.}~\bibnamefont {Wu}},\
  }\href@noop {} {\bibfield  {journal} {\bibinfo  {journal} {Phys. Rev. Lett.}\
  }\textbf {\bibinfo {volume} {94}},\ \bibinfo {pages} {054504} (\bibinfo
  {year} {2005})}\BibitemShut {NoStop}%
\bibitem [{\citenamefont {Jung}\ \emph {et~al.}(2006)\citenamefont {Jung},
  \citenamefont {Mareck}, \citenamefont {Shelley},\ and\ \citenamefont
  {Zhang}}]{jung2006dynamics}%
  \BibitemOpen
  \bibfield  {author} {\bibinfo {author} {\bibfnamefont {S.}~\bibnamefont
  {Jung}}, \bibinfo {author} {\bibfnamefont {K.}~\bibnamefont {Mareck}},
  \bibinfo {author} {\bibfnamefont {M.}~\bibnamefont {Shelley}}, \ and\
  \bibinfo {author} {\bibfnamefont {J.}~\bibnamefont {Zhang}},\ }\href@noop {}
  {\bibfield  {journal} {\bibinfo  {journal} {Phys. Rev. Lett.}\ }\textbf
  {\bibinfo {volume} {97}},\ \bibinfo {pages} {134502} (\bibinfo {year}
  {2006})}\BibitemShut {NoStop}%
\bibitem [{\citenamefont {Jia}\ and\ \citenamefont
  {Yin}(2008)}]{jia2008passive}%
  \BibitemOpen
  \bibfield  {author} {\bibinfo {author} {\bibfnamefont {L.}~\bibnamefont
  {Jia}}\ and\ \bibinfo {author} {\bibfnamefont {X.}~\bibnamefont {Yin}},\
  }\href@noop {} {\bibfield  {journal} {\bibinfo  {journal} {Phys. Rev. Lett.}\
  }\textbf {\bibinfo {volume} {100}},\ \bibinfo {pages} {228104} (\bibinfo
  {year} {2008})}\BibitemShut {NoStop}%
\bibitem [{\citenamefont {Schnipper}\ \emph {et~al.}(2009)\citenamefont
  {Schnipper}, \citenamefont {Andersen},\ and\ \citenamefont
  {Bohr}}]{schnipper2009vortex}%
  \BibitemOpen
  \bibfield  {author} {\bibinfo {author} {\bibfnamefont {T.}~\bibnamefont
  {Schnipper}}, \bibinfo {author} {\bibfnamefont {A.}~\bibnamefont {Andersen}},
  \ and\ \bibinfo {author} {\bibfnamefont {T.}~\bibnamefont {Bohr}},\
  }\href@noop {} {\bibfield  {journal} {\bibinfo  {journal} {J. Fluid Mech.}\
  }\textbf {\bibinfo {volume} {633}},\ \bibinfo {pages} {411} (\bibinfo {year}
  {2009})}\BibitemShut {NoStop}%
\bibitem [{\citenamefont {Rutgers}\ \emph {et~al.}(2001)\citenamefont
  {Rutgers}, \citenamefont {Wu},\ and\ \citenamefont
  {Daniel}}]{rutgers2001conducting}%
  \BibitemOpen
  \bibfield  {author} {\bibinfo {author} {\bibfnamefont {M.}~\bibnamefont
  {Rutgers}}, \bibinfo {author} {\bibfnamefont {X.}~\bibnamefont {Wu}}, \ and\
  \bibinfo {author} {\bibfnamefont {W.}~\bibnamefont {Daniel}},\ }\href@noop {}
  {\bibfield  {journal} {\bibinfo  {journal} {Rev. Sci. Instrum.}\ }\textbf
  {\bibinfo {volume} {72}},\ \bibinfo {pages} {3025} (\bibinfo {year}
  {2001})}\BibitemShut {NoStop}%
\bibitem [{\citenamefont {Jia}\ and\ \citenamefont
  {Yin}(2009)}]{jia2009response}%
  \BibitemOpen
  \bibfield  {author} {\bibinfo {author} {\bibfnamefont {L.-B.}\ \bibnamefont
  {Jia}}\ and\ \bibinfo {author} {\bibfnamefont {X.-Z.}\ \bibnamefont {Yin}},\
  }\href@noop {} {\bibfield  {journal} {\bibinfo  {journal} {Phys. Fluids}\
  }\textbf {\bibinfo {volume} {21}},\ \bibinfo {pages} {101704} (\bibinfo
  {year} {2009})}\BibitemShut {NoStop}%
\bibitem [{exp()}]{experimentcon}%
  \BibitemOpen
  \href@noop {} {\bibinfo  {journal} {The filament is fastened to one end of
  the plate while passes through the hook on the other end and finally wraps
  around a roller behind the film. The hooks are tiny and symmetrically made
  and therefore no obvious impact on the flow is observed. Since fluid forces
  always stretch the filament, rotating the roller smoothly increases or
  decreases $L_d$ accordingly. The roller is rotated gently that the bending of
  cantilever is not influenced. The transverse motion of reflected laser beam
  is not observed}\ }\BibitemShut {NoStop}%
\bibitem [{\citenamefont {Couder}\ \emph {et~al.}(1989)\citenamefont {Couder},
  \citenamefont {Chomaz},\ and\ \citenamefont
  {Rabaud}}]{couder1989hydrodynamics}%
  \BibitemOpen
\bibfield  {journal} {  }\bibfield  {author} {\bibinfo {author} {\bibfnamefont
  {Y.}~\bibnamefont {Couder}}, \bibinfo {author} {\bibfnamefont {J.~M.}\
  \bibnamefont {Chomaz}}, \ and\ \bibinfo {author} {\bibfnamefont
  {M.}~\bibnamefont {Rabaud}},\ }\href@noop {} {\bibfield  {journal} {\bibinfo
  {journal} {Physica D}\ }\textbf {\bibinfo {volume} {37}},\ \bibinfo {pages}
  {384} (\bibinfo {year} {1989})}\BibitemShut {NoStop}%
\bibitem [{\citenamefont {Williamson}(1996)}]{williamson1996vortex}%
  \BibitemOpen
  \bibfield  {author} {\bibinfo {author} {\bibfnamefont {C.~H.~K.}\
  \bibnamefont {Williamson}},\ }\href@noop {} {\bibfield  {journal} {\bibinfo
  {journal} {Annu. Rev. Fluid Mech.}\ }\textbf {\bibinfo {volume} {28}},\
  \bibinfo {pages} {477} (\bibinfo {year} {1996})}\BibitemShut {NoStop}%
\bibitem [{\citenamefont {Sane}\ \emph {et~al.}(2018)\citenamefont {Sane},
  \citenamefont {Mandre},\ and\ \citenamefont {Kim}}]{sane2018surface}%
  \BibitemOpen
  \bibfield  {author} {\bibinfo {author} {\bibfnamefont {A.}~\bibnamefont
  {Sane}}, \bibinfo {author} {\bibfnamefont {S.}~\bibnamefont {Mandre}}, \ and\
  \bibinfo {author} {\bibfnamefont {I.}~\bibnamefont {Kim}},\ }\href@noop {}
  {\bibfield  {journal} {\bibinfo  {journal} {J. Fluid Mech.}\ }\textbf
  {\bibinfo {volume} {841}} (\bibinfo {year} {2018})}\BibitemShut {NoStop}%
\bibitem [{\citenamefont {Roshko}(1955)}]{anatol1955wake}%
  \BibitemOpen
  \bibfield  {author} {\bibinfo {author} {\bibfnamefont {A.}~\bibnamefont
  {Roshko}},\ }\href@noop {} {\bibfield  {journal} {\bibinfo  {journal} {J.
  Aeronaut. Sci.}\ }\textbf {\bibinfo {volume} {22}},\ \bibinfo {pages} {124}
  (\bibinfo {year} {1955})}\BibitemShut {NoStop}%
\bibitem [{\citenamefont {Munson}\ \emph {et~al.}(2013)\citenamefont {Munson},
  \citenamefont {Okiishi}, \citenamefont {Huebsch},\ and\ \citenamefont
  {Rothmayer}}]{munson2013fluid}%
  \BibitemOpen
  \bibfield  {author} {\bibinfo {author} {\bibfnamefont {B.~R.}\ \bibnamefont
  {Munson}}, \bibinfo {author} {\bibfnamefont {T.~H.}\ \bibnamefont {Okiishi}},
  \bibinfo {author} {\bibfnamefont {W.~W.}\ \bibnamefont {Huebsch}}, \ and\
  \bibinfo {author} {\bibfnamefont {A.~P.}\ \bibnamefont {Rothmayer}},\
  }\href@noop {} {\emph {\bibinfo {title} {Fluid mechanics}}}\ (\bibinfo
  {publisher} {Wiley Singapore},\ \bibinfo {year} {2013})\BibitemShut {NoStop}%
\bibitem [{\citenamefont {Bearman}(1967)}]{bearman1967vortex}%
  \BibitemOpen
  \bibfield  {author} {\bibinfo {author} {\bibfnamefont {P.~W.}\ \bibnamefont
  {Bearman}},\ }\href@noop {} {\bibfield  {journal} {\bibinfo  {journal} {J.
  Fluid Mech.}\ }\textbf {\bibinfo {volume} {28}},\ \bibinfo {pages} {625}
  (\bibinfo {year} {1967})}\BibitemShut {NoStop}%
\bibitem [{\citenamefont {Batchelor}(1967)}]{batchelor2000introduction}%
  \BibitemOpen
  \bibfield  {author} {\bibinfo {author} {\bibfnamefont {G.~K.}\ \bibnamefont
  {Batchelor}},\ }\href@noop {} {\emph {\bibinfo {title} {An Introduction to
  Fluid Dynamics}}}\ (\bibinfo  {publisher} {Cambridge Univ. Press},\ \bibinfo
  {year} {1967})\BibitemShut {NoStop}%
\bibitem [{\citenamefont {Yeung}(2009)}]{yeung2009pressure}%
  \BibitemOpen
  \bibfield  {author} {\bibinfo {author} {\bibfnamefont {W.~W.~H.}\
  \bibnamefont {Yeung}},\ }\href@noop {} {\bibfield  {journal} {\bibinfo
  {journal} {J. Fluid Mech.}\ }\textbf {\bibinfo {volume} {622}},\ \bibinfo
  {pages} {321} (\bibinfo {year} {2009})}\BibitemShut {NoStop}%
\bibitem [{\citenamefont {Maybury}\ and\ \citenamefont
  {Rayner}(2001)}]{maybury2001avian}%
  \BibitemOpen
  \bibfield  {author} {\bibinfo {author} {\bibfnamefont {W.~J.}\ \bibnamefont
  {Maybury}}\ and\ \bibinfo {author} {\bibfnamefont {J.~M.~V.}\ \bibnamefont
  {Rayner}},\ }\href@noop {} {\bibfield  {journal} {\bibinfo  {journal} {Proc.
  R. Soc. London. Ser. B}\ }\textbf {\bibinfo {volume} {268}},\ \bibinfo
  {pages} {1405} (\bibinfo {year} {2001})}\BibitemShut {NoStop}%
\bibitem [{\citenamefont {Shoele}\ and\ \citenamefont
  {Zhu}(2010)}]{shoele2010flow}%
  \BibitemOpen
  \bibfield  {author} {\bibinfo {author} {\bibfnamefont {K.}~\bibnamefont
  {Shoele}}\ and\ \bibinfo {author} {\bibfnamefont {Q.}~\bibnamefont {Zhu}},\
  }\href@noop {} {\bibfield  {journal} {\bibinfo  {journal} {J. Fluid Mech.}\
  }\textbf {\bibinfo {volume} {650}},\ \bibinfo {pages} {343} (\bibinfo {year}
  {2010})}\BibitemShut {NoStop}%
\bibitem [{\citenamefont {Kim}\ \emph {et~al.}(2012)\citenamefont {Kim},
  \citenamefont {Huang}, \citenamefont {Shin},\ and\ \citenamefont
  {Sung}}]{kim2012flexible}%
  \BibitemOpen
  \bibfield  {author} {\bibinfo {author} {\bibfnamefont {B.}~\bibnamefont
  {Kim}}, \bibinfo {author} {\bibfnamefont {W.-X.}\ \bibnamefont {Huang}},
  \bibinfo {author} {\bibfnamefont {S.~J.}\ \bibnamefont {Shin}}, \ and\
  \bibinfo {author} {\bibfnamefont {H.~J.}\ \bibnamefont {Sung}},\ }\href@noop
  {} {\bibfield  {journal} {\bibinfo  {journal} {J. Fluid Mech.}\ }\textbf
  {\bibinfo {volume} {707}},\ \bibinfo {pages} {129} (\bibinfo {year}
  {2012})}\BibitemShut {NoStop}%
\end{thebibliography}%


\providecommand{\noopsort}[1]{}\providecommand{\singleletter}[1]{#1}%
%

\end{document}